# The Revenge of Distance: Vulnerability Analysis of Critical Information Infrastructure [1]

Sean P. Gorman, Laurie Schintler, Raj Kulkarni, and Roger Stough
*School of Public Policy, George Mason University, Finley Building, 4400 University Drive, Fairfax, VA 22030, USA*

The events of 9/11 brought an increased focus on security in the United States and specifically the protection of critical infrastructure. Critical infrastructure encompasses a wide array of physical assets such as the electric power grid, telecommunications, oil and gas pipelines, transportation networks and computer data networks. This paper will focus on computer data networks and the spatial implications of their susceptibility to targeted attacks. Utilizing a database of national data carriers, simulations will be run to determine the repercussions of targeted attacks and what the relative merits of different methods of identifying critical nodes are. This analysis will include comparison of current methods employed in vulnerability analysis with spatially constructed methods incorporating regional and distance variables. In addition to vulnerability analysis a method will be proposed to analyze the fusion of physical and logical networks, and will discuss what new avenues this approach reveals. The results of the analysis will be placed in the context of national and regional security and economic impact.

## INTRODUCTION

The events of 9/11 brought a new focus to the vulnerability of the US economy to attack from malevolent forces. As the world globalized and technology advanced in leaps and bounds over the last two decades, new infrastructures emerged – in addition to the roads, rail, and power networks, fiber optic grids connecting computers became vital to a new information economy. The new interconnectedness of the globe and pervasiveness of information technologies, that enabled an economic revolution, are also now the tools that terrorists can use to attack the United States and its economy. The dependence of the new economy on information has made the infrastructures that supply it critical to the functioning and stability of the nation. The new infrastructures of telecommunications and IT are "inherently spatial" (Falk and Abler 1980; Gillespie and Robins 1989). Communication systems that lie at the heart of telecommunications and IT compress time and space, reducing, if not eliminating, the effects of distance (Atkinson 1998; Brunn and Leinbach 1991; Cairncross 1997; Castells 1989; Negroponte 1995). The "death of distance" (Cairncross 1997) has led to wide speculation across many disciplines that the IT and telecommunications revolution would be an end to the "tyranny of geography" (Gillespie and Robins 1989).

[1] The authors would like to thank Critical Infrastructure Protection Project at the George Mason School of Law for their funding of the research. The authors would also like to thank Dr. Ed Malecki for loaning data from his Internet infrastructure research - National Science Foundation grant BCS-9911222 and Lei Ding for his assistance with data compilation.

This disassociation of location has led to the common conception that the Internet and IT are virtual entities residing in cyberspace. This same conception has resulted in a belief that security issues for the Internet and IT reside solely in cyberspace as well. While cyber-security concerns such as denial of service attacks, identity theft, and various other forms of hacking are serious security threats, they are not the only danger to the US information infrastructure. The Internet and IT depend on physical fiber to connect the various computers, servers, switches, and routers that provide the underpinnings of the US information infrastructure. All of these vital components have a physical location, but since the US information infrastructure is privately owned and proprietary these locations are most often undisclosed. As a result there is no current map of the US information infrastructure (Internetweek 2001). Without an aggregated network to map there is no process by which to determine if the network is susceptible to a targeted physical attack, and if so what nodes and links are most vital.

Current security mandates state that the telecommunications infrastructure is densely connected and that physical attack is unlikely to disrupt the network for extended periods of time (NRC 2002, NSTAC 2002). While this assertion could be largely true for any one individual network, the susceptibility to attack of the interconnection of the mass of networks comprising the information infrastructure remains to be validated. Empirical evidence has illustrated a high order of attack susceptibility for the interconnected router infrastructure of the Internet (Albert et al 2000, Callaway et al 2000, Cohen et al 2001). The average performance of the Internet would be cut in half if just 1% of the most highly connected routers were incapacitated and loses its integrity with 4% of the most connected routers destroyed. Where are these top 1% and top 4% of routers? Are they distributed enough that a coordinated attack would be infeasible? Are the back up systems and redundancy of private providers sufficient to compensate for these susceptibilities? These are just the first order of questions about the susceptibility of critical infrastructure to attack. Further areas of investigation include the dependencies of other critical infrastructures (finance and banking, electric power etc.) on information networks. Also how can cyber-attacks be used in conjunction with attacks on physical infrastructure to compound damage for greater ripple effects in the economy?

This study is intended to begin to explore some of these issues. Through mathematical and spatial analyses the study will attempt to gain a better understanding of the topology and structure of our nation's complex telecommunications infrastructure – i.e., where the "critical" nodes and links are located in this network, and how the performance of the network is affected by the removal of certain nodes or links. The findings of this analysis will be used to derive a set of policy and planning recommendations on how best to mitigate the catastrophic and cascading effects that could occur as the result of a targeted physical and/or cyber attack on the nation's telecommunications infrastructure. In order to fully understand the implications of the issues addressed in this study and to identify related research efforts across various disciplines, the relevant literatures are next discussed. This survey will briefly cover research of geographic networks and complexity research involving networks. Following the literature review a methodology will be presented for vulnerability analysis of information networks and also an approach to fusing these operational networks with physical networks they run over for analysis. Lastly, the results of the analysis will be placed in the context of national and regional security and economic impact.

## GEOGRAPHIC NETWORKS AND INFORMATION

The most prominent aspect of the telecommunications and information revolution is its ability, through phenomena like the Internet and satellite transmissions; to connect geographically separated locations (that posses the appropriate technology levels). There has been an increased attention of the spatial aspects of communications, but they have often overlooked the foundations of the spatial analysis of networks. Social science and specifically geography has a deep tradition in network and spatial analysis. Garrison and Marble (1962) did in-depth network analysis on the interstate highway system, analyzing the importance of nodes and links on location and development. This same vein of research was greatly expanded through Garrison's student Kansky (1963) and later with the work of Chorley and Haggett (1968). In addition, Nyusten and Dacey (1968) and later Taafee and Gauthier (1973) expanded this research, applying network analysis to telephone networks and general infrastructure. This tradition of network analysis was picked up again by geographers to begin to analyze the Internet's network of networks.

Specifically, examination of the backbone level of the Internet is an area that has enjoyed a significant level of analysis by geographers, planners, policy analysts, and regional economists (Wheeler and O'Kelly 1999, Moss and Townsend 2000, Gorman and Malecki 2000, Malecki and Gorman 2001, Grubesic and O'Kelly 2002, Townsend 2001, Choi et al 2001, Pelletiere and Rodrigo 2001). The findings of this research have been remarkably consistent. Backbone networks disproportionately agglomerate in the largest metropolitan areas, "The critical importance of access to new technologies has highlighted the characteristic diffusion pattern: hierarchical – beginning first in large cities, where the largest markets are found, and then to progressively smaller places (Malecki 1999)". These large markets cluster spatially into distinct groups like the Boston-Washington corridor and the LA-San Francisco complex that are fully interconnected to each other both by the telephone network (Langdale 1983) and the Internet backbones (Gorman and Malecki 1999; Moss and Townsend 1998; Wheeler and O'Kelly 1999). The result is that the spatial hierarchy of the United States is unique; the coasts are more connected to each other than they are to the interior of the country – the coasts are the core and the interior is the periphery (Gorman and Malecki 2000).

Statistical analysis has produced a very distinct profile for these highly connected hub cities. Leading cities in advanced telecommunication services have been found to have high concentrations of producer services, financial service (FIRE), low levels of manufacturing, high population, and high total personal income (McIntee 2001, Malecki 2000, Gorman and McIntee 2001, Leyshon 1996, Warf 1995, Graham 1999). Further, these cities are predominantly world cities that serve as hubs for information services as well as serving a command and control functions for global firms (Know and Taylor 1995, Friedmann and Wolf 1982, Townsend 2001, Malecki 2002).

The profile of these super-connected cities also influences the structure of the network at a micro level within the city. Longcore and Rees' (1996) examination of New York City's financial district demonstrates meaningful analysis of information technology and networks at a very local level. They found that in New York the historic Wall Street-centered financial district has had to disperse within the city to take advantage of buildings and real estate with sufficient built-in infrastructure and technology to handle the district's specialized role in the information economy. Without buildings lit by fiber optic connections financial district firms were disconnected from the global economy.

Network analysis of the spatial networks of the Internet has only begun to be investigated. Wheeler and O'Kelly (1999) examined the basic graph measures of several domestic US providers and analysis of city connectivity of the aggregated providers. Gorman and Malecki (2000) investigated the network topologies of several firms and how graph theoretic measures could be used to investigate competitive advantage and the nature of interconnection between networks. Later studies have looked at the structure of networks and city connectivity as a time series finding large changes in bandwidth capacity (Malecki 2002, Townsend 2001), but little change in graph measures of connectivity (O'Kelly and Grubesic 2002). While connectivity indices have changed little over time the overall structure of the network has – Gorman (2001) found that the aggregated US backbone network has increasing self-organized from 1997 to 2000 creating a more efficient but more sparsely connected network. This research confirmed at a spatial level of analysis what was being found at a topological level in the study of complex networks.

**COMPLEXITY OF TELECOMMUNICATION NETWORKS**

Telecommunications networks are large and complex, and the study of complex networks has enjoyed a flurry of new findings in the last few years. The basis on studying large networks is built upon the work of Erdos and Renyi (ER). Erdos and Renyi (ER) endeavored to use "probabilistic methods" to solve problems in graph theory where a large number of nodes where involved (Albert and Barabasi 2002, p. 54). Under this assumption they modeled large graphs utilizing algorithms where $N$ nodes were randomly connected according to probability $P$, and found that when vertices were connected in this fashion they followed a Poisson distribution (Albert and Barabasi 2002, p 49). A more thorough review of random graphs can be found in the survey work of Bollobas (1985).

The absence of detailed topological data for complex networks left random network models as the most widely used method of network simulation (Barabasi 2001).

Watts and Strogatz (WS)(1998) using several large data sets found that complex networks were not entirely random but instead displayed significant clustering at the local level. Further, that local clusters linked across the graph to each other forming "small worlds". These short cuts across the graph to different clusters of vertices introduced a level of efficiency not predicted in the ER model and showed the first signs of self-organization in complex networks. The distribution was not Poisson as with the ER model, but was bounded and decayed exponentially for large sets of vertices. These findings spurred a flurry of work into the attributes of complex networks and new findings and discoveries quickly followed. In 1998 two parallel studies by Barabasi, Reka, and Jeong of Notre Dame and Huberman and Adamic at Parc Xerox found that when one looks at the World Wide Web as a graph (web pages are vertices and hyperlinks connecting them are edges) it followed not a Poisson or exponential distribution, but a power law distribution.

A power law is a significantly different finding from either the expected exponential or Poisson distribution. In a power law distribution there is an abundance of nodes with only a few links, and a small but significant minority that have very large number of links (Barabasi 2001). This is a distinct difference from both the ER and WS model that should be noted; the probability of finding a highly connected vertex in the ER and WS model decreases exponentially, thus vertices with high connectivity are practically absent (Barabasi and Albert 1999). The reason, according to Barabasi and Albert, was that complex networks were not static; the number of nodes does not stay constant as in the WS and ER model. Complex networks grow over time and new vertices attach preferentially to already well-connected vertices in the network

Barabasi and Albert (1999) formalized this idea in "Emergence of Scaling in Random Networks". They stated that in a complex network like the World Wide Web the probability *P(k)* that a vertex in the network interacts with *k* other vertices decays as a power law following $P(k) \sim k^{-g}$. Further, that this result indicates that large networks self organize into a scale-free state, a feature unpredicted by all existing random network models. The Barabasi-Albert (BA) model is based on three mechanisms that drive the evolution of graph structures over time to produce power law relationships:

1. Incremental growth – Incremental growth follows from the observation that most networks develop over time by adding new nodes and new links to existing graph structure.
2. Preferential connectivity – Preferential connectivity expresses the frequently encountered phenomenon that there is higher probability for a new or existing node to connect or reconnect to a node that already has a large number of links (i.e. high vertex degree) than there is to (re)connect to a low degree vertex.
3. Re-wiring – Re-wiring allows for some additional flexibility in the formation of networks by removing links connected to certain nodes and replacing them with new links in a way that effectively amounts to a local type of re-shuffling connection based of preferential attachment.

This difference between the ER and BA model become clear when seen in a visual representation, Figure 1 illustrates the structural difference between a random ER network model and a scale free network model. The high level of clustering and super-connected node is evident on the node diagram. For the model in Figure 1, "more than 60% of nodes (green) can be reached from the five most connected nodes (red) compared with only 27% in the random network. This demonstrates the key role that hubs play in the scale-free network. Both networks contain the 130 nodes and 430 links" (Barabasi 2001 p.3). The implications of this were very broad for a number of disciplines as varied as genetics, economics, molecular physics and sociology. One of the surprising findings was that not only did the World Wide Web fall into a scale free organization, but so did the Internet.

The Faloutsos brothers (1999) found that the Internet followed power laws at both the router level and autonomous system (AS) level. The router level entails the fiber optic lines (edges) and the routers (vertices) that direct traffic on the Internet, and the AS level entail networks (AT&T, UUNet, C&W etc.) as vertices and their interconnection as edges. This meant that that the physical fabric of the Internet and the business interconnections of the networks that comprise the Internet both qualified as scale free networks. Before these discoveries, the Internet had been modeled as a hierarchy and the new finding had many implications throughout the field of computer science. The scale free theory and BA model have not been without debate. Several arguments have been made stating that the BA model and is too simplistic for the Internet and additional corollaries need to be made (Chen et al 2001). The re-wiring principle was one response to these criticisms, but overall the model has held (Albert and Barabasi 2000). Tests of network generators based on power laws have been found to produce better models and efforts are being made to base new Internet protocols on these discoveries (Tangmunarunkit et al 2001, Radoslavov et al 2001). While these discoveries have paved the way for advancements in several fields the question of the geography and location of these networks remain to be addressed.

**ERROR AND ATTACK TOLERANCE OF COMPLEX NETWORKS**

Scale free networks have many implications, but a far-reaching consequence of their unique structure is they are very fault tolerant but also susceptible to attack (Albert et al 2000). Research done at the Notre Dame Center on Self Organizing Networks found that a scale free network model remains connected when up to 80% of nodes are randomly removed from the network. On the other hand, when the most connected nodes are removed the diameter of the network increases rapidly, doubling its original value if the top 5% of nodes are removed. This work was confirmed by the analysis of Callaway et al (2000) modeling network robustness and fragility as a percolation and by Cohen et al (2001). This work was extended to demonstrate that scale-free dynamical networks have the same robust and fragile properties under synchronization. Preliminary analyses of these models on spatial network data have shown similar results when cities are the nodes and fiber connections between them are the links (Gorman and Kulkarni 2002).

Utilizing a different model of node connectivity and path availability Grubesic et al (forthcoming) find that the disconnection of a major hub city can cause the disconnection of peripheral cities from the network. Spatial analysis of network failure has also been done for airline networks finding similar results for the Indian airline network (Cliff et al 1979). These findings have profound implications for telecommunications networks that have scale free properties.

## METHODOLOGY

In order to test the susceptibility of the spatial US information infrastructure to attack several approaches were developed building upon prior studies and literature. Several studies have noted that the removal of critical nodes caused a rapid degradation of the network. Generally, these nodes are determined to be critical based on their level of connectivity, often referred to as an accessibility index in spatial network analysis. The first methodological step of this study was to determine which nodes are most critical and what is the best way to identify these nodes. The next step is to test the set of nodal hierarchies and see which ranking method has the greatest impact on the network when nodes are targeted for failure and what the ripple effect of their removal is. Lastly, an attempt will be made develop analysis that will allow the modeling of physical fiber networks with logical (operational) networks. This approach will allow some understanding of the impact a cut or failure in the physical network will have on the logical network. Data will be analyzed for each step of the methodology and conclusions will be drawn.

*Nodal Hierarchies*

There are several approaches to establishing nodal hierarchies within a network, and classical spatial network analysis has some standard approaches. The most common of which is the accessibility index:

$$A_i = \sum_{i=1}^{n} d_{ij} \qquad (1)$$

The accessibility index provides the number of connections to a node; in the case of this study each MSA is considered a node. Another derivation of the accessibility index is to look at not only the binary connectivity of a node, but also the capacity of those links:

$$A_i^{'} = \sum_{i=1}^{n} c_{ij} \qquad (2)$$

*c = capacity of the link*

While both of these approaches to nodal hierarchies provide useful insight into connectivity there are some drawbacks. These nodal hierarchies are based on some fundamental assumptions:

1. There are a certain number of discrete-classes of settlements.
2. The number of settlements in each size class increases down the hierarchy.

3. There is some characteristic spacing between nearest neighbors at any particular level.

(Lowe and Moryadas 1975)

The drawback of this approach is that it is based on a spatial-hierarchical structuring of an idealized Loschian landscape. In a Loschian landscape nodes will connect and goods will flow based on proximity with larger places connecting to each other through intervening smaller places (Hagget 1966). This approach is based on a planar network, where the intersection of any two links results in a node. Information networks, though, are non-planar constructs where two distant places can be directly connected without intervening nodes being transited. As a result traditional methods of determining nodal hierarchies might not be appropriate.

Recent research into complex networks has found that information networks often form small world or scale free networks. The general attribute of these formations is that there are formations of local clusters that are interconnected through global connections. This creates a highly efficient and sparsely connected network. Vulnerability studies have focused on identifying the global connections in these systems and then targeting them for failure or attack (Albert et al 2000, Cohen et al 2001, Callaway et al 2001). These studies all found that removing the global connectors in the network led to catastrophic failure. This leaves the question of how global connectors should be identified in a spatial network? Further, how should local clustering and global connections be determined?

One approach is to use regions to define what is global and what is local (Gorman and Kulkarni 2002). In this approach the United States is divided into the four census regions (South, West, Midwest, Northeast) and each city in the information infrastructure matrix was assigned to a census region. For each city the number of local links to other cities in the same census region were totaled along with the number of global links connecting to cities in other census regions. From this data the following approach was developed to identify cities that act as the super connected cities that provide the key global connection in the network:

Consider a large network of *n* nodes, spanning an area *A* consisting of *m* regions, with variable number of nodes inside each region that have variable number of connections from each region to other regions. For a region *r* with *p* number of nodes, a *pxp* contiguity matrix represents connections between these nodes. Then, one could construct a contiguity or adjacency matrix for the entire network of *m* regions, as a block diagonal matrix, where matrices along the main diagonal refer to the contiguity matrices for each of the regions, while interregional connections are represented as the off-block-diagonal elements. Let *M* denote such a matrix.

If a node *i* in region *r* is connected to another node *j* in the same region, then that connection is considered as a local link and is denoted by $q_{i(r)j(r)}$. On the other hand if node *i* in region *r* is connected to node *k* in region *s* then that connection is considered as a global connection and is denoted by $g_{i(r)k(s)}$. Thus, in theory, one may associate with each node node *i(r)*, a global connectivity index as a ratio between its global and local connections, weighted by the total number of global and local connections for the entire network.

The total number of global connections G are computed from the elements of the block upper triungular matrix of M, of $m$ regions, each with variable number of nodes:

$$G = \sum_{i(1)} \sum_{s>1}^{m} \sum_{k(s)} g_{i(1)k(s)} + \sum_{i(2)} \sum_{s>2}^{m} \sum_{k(s)} g_{i(2)k(s)} + \ldots + \sum_{i(m-1)} \sum_{s>m-1}^{m} \sum_{k(s)} g_{i(m-1)k(s)} \qquad (1)$$

Note that, since $m$ is the last region in the block diagonal matrix, its global connections have already been computed in the previous $m$-1 blocks.

The total number of local connections L is a sum over all the local connections over $m$ regions and is given by:

$$L = \sum_{i(1)} \sum_{j(1)>i(1)} q_{i(1)j(1)} + \sum_{i(2)} \sum_{j(2)>i(2)} q_{i(2)j(2)} + \ldots + \sum_{i(m)} \sum_{j(m)>i(m)} q_{i(m)j(m)} \qquad (2)$$

Then the global connectivity index for a node $i$ in region $r$ is given by:

$$C_{i(r)} = \left( \frac{\sum_{s \neq r}^{m} \sum_{k(s)} g_{i(r)k(s)}}{1 + \sum_{j(r), j \neq i} q_{i(r)j(r)}} \right) \times (G + L) \qquad (3)$$

The numeral of 1 in the denominator indicates a self-loop of a node. The ratio of global links to local links provides an indicator of how well the city acts as a global connector in the network and the weighting of the scores by the total number of links balances the measure with the overall connectivity of the node.

This approach does have several drawbacks, most obvious being a border effect. Those nodes located close to a border can have a very short link that crosses to another census region, artificially inflating the number of global connections. An alternative approach is to base the global-local assignment with the Euclidean length of the link, a distance based small world. Any link over *x* miles is considered global and anything under *x* miles is considered local. This approach would yield the following equation:

$$\left( \frac{\sum_j g_{ij} > D}{1 + \sum_j l_{ij} \leq D} \right) \left( \sum_j g_{ij} + \sum_j l_{ij} \right) \qquad (6)$$

The problem lies in what distance should be the cut off for what is considered global or local. To gain some perspective on this problem the equation was automated, and a series of distances for *D* were simulated:

$$\text{Where} \quad D \in [100, 200, 300 \ldots 2700] \qquad (7)$$

The simulations produced the graph seen in figure 2, where the x-axis is the increments of *D* tested and the y-axis is the percentage of nodes with a global to local ration greater than one. Figure 2 shows a sharp shift somewhere before 500 miles. To find the exact point of inflection the rate of change in the global to local ration was calculated, as illustrated in figure 3. The rate of change graph clearly points to 300 miles as being the point of inflection. Under such an assumption all links shorter than 300 miles are considered local and all links over 300 miles are considered global. Interestingly 300 miles is very close to the mean distance of all links in the matrix, 308.6 miles. Further, the average length of private leased lines (a different set of data) in the United States is 300 miles (Coffman and Odlyzko 1999). The recurring nature of 300 miles could point to a common spatial structure in the United States for communication networks. This is an area that merits further investigation and possible cross-country comparisons.

Another hierarchy was developed based on the distance small world approach looks at just the number of global connections a node has, and ranks them based on that count. The equation would then be:

$$\sum_j g_{ij} > D \qquad (6)$$

This ranking would provide an indicator of how many long haul global connections a city has. The hypothesis being that the more global connections the greater the impact the loss of the node will have in connecting various clusters across the network. A final approach considered for this study is to identify relay nodes and what effect they have on the survivability of the network. Relay nodes are typically defined as locations that are neither the ultimate origin nor destination of an interaction across a network.

The primary purpose of a relay node is to receive flows in order to transmit them to another node with minimum delay and cost (Lowe and Moryadas 1975). On the information networks like the Internet the definition of relay node is more fluid. Nodes constantly shift from being origin, destination, or relay nodes. This requires sorting of which nodes are disproportionately used as relay nodes. Nodes that act as structural links to relay information to large markets could serve as critical junctures. To sort out which nodes were disproportionately acting as relay nodes the following approach was developed

$$\frac{\sum_{i=1}^{n} c_{ij}}{\sum_{i=1}^{n} b_{ij}} \qquad (7)$$

Where $c_{ij}$ = capacity and $b_{ij}$ = business demand

This approach provides a rough indicator of how much built capacity exceeds the consumption of capacity dictated by demand. It is assumed that nodes with an aggregate disproportionate amount of excess capacity are using it to relay information to other destination nodes. The hypothesis for a relay node hierarchy is that these nodes are structurally important in linking up large destination nodes and without them connection would be severely affected.

With a series of nodal ranking approaches designed, a data set is needed to apply them to. For this study a 2000 dataset of aggregated IP network providers was used comprising of a matrix of 147 metropolitan statistical areas and the IP bandwidth available between each one (Malecki 2001). It should be noted that the accuracy of the data is not perfect. IP network provider's maps often advertise more capacity than is currently in operation and future routes are often shown as current routes. With the data accuracy shortcoming in mind the data does provide an adequate test base to compare algorithms and examine rough rankings. Further, the aggregation of this dataset assumes that all networks interconnect with all other networks in every city they collocate in. This is far from the case in reality, but provides a best-case scenario for testing purposes. A more realistic model is being developed as a future direction of this study. From this capacity matrix, a binary connectivity matrix and Euclidean distance matrix were constructed. Finally business demand for the relay node hierarchy was calculated from FCC figures tabulated by Telegeography (2002). The results of each hierarchy (accessibility, aggregate capacity, spatial small world, distance small world, and relay node) can be found in table 1 as part of the analysis results.

In order to test these nodal hierarchies with the datasets outlined above each rank-order was run through two simulations. In the first simulation each node was successively removed according to its rank and the diameter of the network was measured for each removal. The diameter of the network is the minimum number of hops it takes to get from the two furthest nodes on the network. Mathematically this is expressed as:

Diameter = maximum $D_{ij}$       $D_{ij}$ = shortest path (in links) between

the *i*th and *j*th node

The same simulation was run again, except this time instead of monitoring the diameter of the network the S-I index was examined. The S-I index of a graph is based on the frequency distribution of the shortest path lengths $s_{ij}$ in the graph. It is defined as the pair *(S,I)*, where:

$$S = \frac{m_3}{m_2} \quad and \quad I = \frac{m_2}{m_1'} \qquad (8)$$

According the Cliff et al (1979 p. 45) "The values of the *S* and *I* can be calculated for a variety of theoretical distributions based on the hypergeometric series, and can be mapped on the *S-I* plane." The resulting S-I plane forms a hoop with one end indicating a fully meshed (interconnected network) and the other a minimally spanning tree. By examining the S-I index of the US IP network infrastructure as nodes are removed one can get a quantitative indication of how disconnected the network becomes.

The results of both the diameter and S-I index analysis can be found in table 2. The diameter results are the easiest to interpret and reveal some interesting findings. The hierarchies with the largest effect on the diameter of the network were the small world distance hierarchy and the global hierarchy, both of which ended is a diameter of 16 when the top 15 nodes (roughly 10%) were removed. The starting diameter of the network in each case was 7 and the result of 16 was more than a doubling of the diameter, meaning it took more than twice the number of hops to reach the two furthest places on the network. This results in a ripple effect across the network where it will take a minimum of twice the time to get from any point to another. This does not take into account the capacity of the links removed and how traffic will be redistributed across the network. This is an area of ongoing research based on this study. While both hierarchies end up at 16 the global hierarchy accelerates more rapidly in the beginning while the small world distance hierarchy accelerates the diameter more quickly at the end of the nodal hierarchy. The next group of nodal hierarchies was the relay node and small world regional hierarchy which both end with a diameter of 12. Finally, the binary and bandwidth capacity hierarchy had the least impact each ending in a diameter of 10.

The diameter relationship of the hierarchies is seen more clearly when all the nodal hierarchies are plotted with their diameters at each successive node removal (figure 4). The lower impact of the binary and bandwidth hierarchy are very evident up to roughly the fifteenth node removal, when the diameter starts to vary erratically first for small world distance then global and finally small world regional. The decrease in diameter results from the network Balkanizing into two or more sub-graphs. The sub-graph is a smaller subset of the whole network and thus has a lower diameter. This is a significant juncture because individual places are no longer just being disconnected, but entire islands of nodes are being created, signifying a catastrophic failure of the network. Further, the bottom placement of the binary connectivity hierarchy is of note since this is the method by which nodes have been chosen for removal in the majority of studies in the literature (Albert et al 2000, Callaway et al 2000, Cohen et al 2001). This would appear to indicate that the most critical nodes in the network are not the most obvious ones, and judging nodes by only the agglomeration of links is not a sufficient method for examining the susceptibility of spatial networks. The study does reveal that incorporating distance into an analysis of critical nodes for network survivability produces significantly improved results. Finally, when it comes to the survivability of the Internet distance is hardly dead.

An examination of the S-I index confirms the findings of the diameter analysis. Figure 4 illustrates the S and I measure of the network as nodes are removed from the network, specifically the global hierarchy approach. The graph clearly shows the similar effect S and I have with diameter as nodes are removed and the extreme sensitivity of S to network changes. The graphical approach is different from the typical plotting of the S and I onto the S-I plane as (X,Y) coordinates, but works well in this case to one demonstrate the connection between diameter and the S-I measures, and two show how increases in the S-I index are indicators of a disconnecting network.

*PHYSICAL DISJOINT ANALYSIS*

The methodology and approaches outlined thus far have only dealt with one facet of US infrastructure, networks that operate with Internet protocol (IP). These IP networks are operational constructs and their topology is logical not physical. The nodes in an IP network are physical places but they way they are connected are not. The map of a typical IP network will show direct connections between far-flung places like New

York and San Francisco or Seattle and Chicago. The logical connections of IP networks are carried over physical fiber in the ground, which form much different network constructs. Physical fiber networks typically run over rights of way established by physical transportation networks – roads, railroads, pipelines, or sewers. Fiber is laid connecting several cities and then either leased to provide connectivity to other networks or to run the networks of the fiber provider. These operational networks then choose from a variety of protocols and technologies to set up an operational network to connect their key assets. The most popular current protocol is IP and the most popular network is the Internet, but there are wide variety of other protocols and networks that utilize the leased lines of physical fiber networks. A wide variety of critical infrastructures depend on leased lines for their operation, ranging from banking and finance to military command and control. The aggregate private leased lines in the United States dwarf the bandwidth available on the Internet – in 1997 leased lines accounted for 330 Gbps and the Internet only 75 Gbps (Coffman and Odlyzko 1999). . This relationship has only intensified with the telecom boom and bust:

> with very few exceptions all ISPs have typically just a single OC48 or at most OC192 link along their major routes. Yet the facilities based carriers typically have between 40 and 800 fibers along each route, and each fiber is usually capable (with current DWDM technology) of carrying 80 OC48 or OC192 wavelengths. Thus only a small fraction of the fiber capacity is currently used for Internet traffic (Odlyzko 2002 p.5).

The vast majority of this capacity remains unused or is allocated to private leased lines. In fact, if you take the top twenty US cities and average how much lit bandwidth is used for the Internet it is only 1.37% of the total (based on Telegeography 2002). While the Internet is a small percentage of total capacity it is a much larger percentage of traffic averaging 2,500-4000 TB a month compared to 3,000-5,000 TB a month for private lines (Coffman and Odlyzko 1999). As such the Internet is a good indicator of a very active operational network.

The often-confusing part is that the Internet is collection of interconnected private lines and networks. The difference between the categories outlined by Coffman and Oldyzko (1999) is that the Internet all runs on TCP/IP and the individual private networks (autonomous systems) agree to interconnect with each other under a common framework. Where as private lines, in their classification, are not openly accessible for interconnection under a common framework and protocol. This still leaves the problem of how does one analyze a network that has two components, one physical and one logical with two different topologies. One approach is physical disjoint analysis developed by Bhandari (1999), where a logical network is overlaid on the physical network it runs over. In this graph theoretical construct a new terminology is added, spans and span nodes. Spans are the actual physical links that comprise the network and logical links are built from these spans. Thus, a given span can be common to a number of links, and several spans can combine to form a single logical link (Bhandari 1999). Once a physical and logical network is combined there are logical links and nodes and span links and nodes. This approach makes more sense when illustrated visually. Figure 6 is the physical fiber network for Genuity and Figure 7 is the logical IP network that runs over the physical fiber network. Figure 8 illustrates the combination of the physical fiber and operational IP network into one map.

The goal of physical disjoint analysis is to examine these two network as graphs and determine how many physically disjoint paths exist between any two places on the network. Thus, if there were a physical fiber cut would there be another physical path for the data to follow to arrive at its destination. Mathematically the edge disjointness (ED) for a pair of paths can be defined as:

$$ED = 1 - \frac{\sum l_j}{\sum l_i} \qquad (9)$$

Where the $j$ sum is over the common edges of the two paths and the $i$ sum is over the edges of the two paths. This approach becomes clearer when framed as a visual example. In the Genuity example above how many physically disjoint paths are there between San Francisco and New York? While there is a direct connection between San Francisco and New York in the logical network it requires a minimum of 7 physical links to connect the two. There is also a second route that does not use any links in the 7-link path, comprising a completely physically disjoint path with 11 links. Thus its ED would be 1 since there are no common edges shared between the two. If there was a shared path then the ED would be 1 – (1/11+7) = .944. While this approach provides a good tool for examining specific routes it does not give any insight to the system dynamics of the network or the criticality of links in a route. Building upon the ED approach the two physically disjoint paths illustrated between San Francisco and New York were analyzed to see what the impact of two strategic cuts would have on the logical network. One cut was made between Denver and Kansas on the 7-link disjoint path and the other cut between El Paso and Houston on the 11-link route. Cut one results in the loss of 8 logical connections and cut two results in the loss of two logical connections and the disconnection of the East and West coast. This provides only the characteristic for the routes between two nodes. In order to examine how critical these two links are for the entire network a link frequency analysis was constructed. This approach analyzes the frequency for which a particular link must be traversed for each iteration of a network until the diameter of the network is reached. Thus the first iteration would examine how many times link $x$ would have to be traversed in all combinations of two hop routes, and the third iteration would be the percentage of times a link had to be traversed for all three hop combinations and so on, till the diameter was reached. When this analysis was run for the two links under examination in the Genuity fiber network the results seen in table 3 were produced. The results show that for all two-hop combinations in the network the two links are utilized in 15% of routes. This number increases for each iteration of the network until 97% of all routes utilize these two link for all 7-hop combinations and 100% of all 8-hop combinations are dependent on these two links. This approach can also be used to rank the relative importance of links of the network and calculate how many logical connections are dependent on them as illustrated earlier.

**CONCLUSIONS**

Identifying and determining the location of vital nodes in US critical infrastructure has taken on renewed importance since the events of 9/11. The critical infrastructure of today's economy has grown from transport and utilities to include information networks like the Internet, telecommunications, and various financial networks. These networks are structurally different from traditional infrastructure networks, and this paper has endeavored to develop new methods of identifying critical nodes and links in them. The new approaches developed had a far greater impact on the connectivity of the network than traditional approaches. Interestingly distance based approaches (small world distance model and global hierarchy model) had the greatest impact on the network, and outperformed traditional measures (accessibility and capacity models) by 62.5%. These results reveal the important role distance plays in the structural properties of small world and scale free networks, especially in the case on information infrastructure networks. Also of note is the reoccurring importance of 300 miles as a threshold in distance analysis of information networks. It was the critical inflection point between global and local in this study, the average length of leased lines in the Coffman and Odlyzko (1999) study, and the average logical link length in this study. It would appear, that in the case of vulnerability, distance has avenged the premature calls of its demise (Cairncross 1999).

This analysis has also put forward a preliminary method of analyzing the interactions of physical and logical networks. Utilizing the concept of spans from Bhandari (1999) it is possible perform physical disjoint analysis and determine the impact of physical fiber cuts on a logical network. The approach was extended to allow an overall system analysis to determine the criticality of physical links to the operation of a logical network. This allows vulnerability analysis of operational networks and the private line infrastructure they are dependent on at both a micro and macro level. This is a first step in analyzing the interdependencies of various information networks (financial, military, aviation, energy) and the fiber infrastructure they are dependent on.

Applying the analysis of this paper to create public policy recommendations proves to be a difficult task. Unlike traditional infrastructures fiber networks are almost entirely held by private firms. Multiple providers supply the infrastructure, often interconnecting with each other and always in tight competition, creating unique interdependencies. One network's security is only as good as the other networks it interconnects with (Kunreuther et al 2002). Thus there is not a direct economic incentive to secure a network if it can be compromised by the competition, a classic case of prisoner's dilemma. In a telecommunications industry experiencing its largest economic downturn in history the incentive for security is only further pushed to the rear. This raises the question, is a policy intervention needed if the market is not fulfilling the role of providing adequate security measures? This question goes outside of the scope of this paper, but the analysis does point out some fundamental recommendations that should be fulfilled by the market or a policy intervention:

1. Technological and market forces have reduced available reserve capacity and the number of geographically diverse routing paths.

2. Failure of single links and nodes can have serious repercussions and having back up providers is not always a safe solution

3. Operational networks need to be mapped to physical networks to determine susceptibilities.

The best way to secure information networks to cyber and physical attack remains an open question with large implications for the US economy and national security.

**BIBLIOGRAPHY**

Albert, R., Jeong, H., and Barabasi, A. (1999) The diameter of the World Wide Web. *Nature* 401: 130-131.

Albert, R., Jeong, H. and Barabási, A.L. (2000) Attack and error tolerance in complex networks. *Nature* 406: 378

Albert, R. and Barabasi, A. (2001) Statistical mechanics of complex networks. Submitted for publication.

Atkinson, R. (1998) Technological Change and Cities. *Cityscape: A Journal of Policy Development and Research* 3: 129-171.

Barabasi, A. (2001) The physics of the Web. *Physics World* July 2001. http://www.physicsweb.org/article/world/14/7/09

Barabasi, A. and Albert, A. (1999) Emergence of scaling in random networks. *Science* Oct: 509-512.

Bhandari, R. (1999) *Survivable Networks: Algorithms for Diverse Routing.* Boston: Kluwer Academic Press.

Brunn, S D. and Leinbach, T R. (eds) (1991) *Collapsing Space & Time: Geographic Aspects of Communication & Information.* New York: Harper Collins Academic.

Cairncross, F. (1997) *The Death of Distance.* Boston: Harvard Business School Press.

Callaway, D.S., Newman, M.E.J., Strogatz, S.H. and Watts, D.J. (2000) Network robustness and fragility: percolation on random graphs. *Physical Review Letters* 85:25

Castells, M. (1989) *The Information City.* Oxford, UK: Blackwell.

Chen, Q., Hyunseok, C., Govindan, R. Sugih, J. Schenker, S. and Willinger, W. (2001) The origin of power laws in Internet topologies revisited. Submitted for publication.

Choi, J., Chon, B., Barnett, G. and Choi, Y. (2001) World cities and Internet backbone networks: A network approach. presented at Digital Communities 2001: Cities in the Information Society November 4, 2001.

Cliff A., Haggett P., Ord K. (1979) *Graph theory and geography*. <u>in</u> Wilson R., Beineke L. (Eds) Applications of graph theory. Academic Press, London.

Cohen, R., Erez, K., ben-Avraham, D., and Havlin, S. (2001) Breakdown of the Internet under intentional attack. *Physical Review Letters* 86:16

Falk, T., Abler, R. (1980) Intercommunications, Distance, and Geographical Theory. *Geografiska Annaler,* Series B, 62: 35-56.

Faloutsos, C., Faloutsos, P., and Faloutsos, M. (1999) On power-law relationships of the Internet Topology. in *Proceedings of the ACM SIGCOMM* Sept.

Garrison, W. (1968) Connectivity of the interstate highway system. In Berry, B. and Marble, D. (1968) *Spatial Analysis.* Englewood Cliffs, NJ: Prentice Hall pp. 239-249.

Gillespie, A. and Robins, K. (1989) Geographical Inequalities: The Spatial Bias of the New Communications Technologies. *Journal of Communications* 39 (3): 7-18.

Gorman, S.P. and Maleci E.J. (2002) Fixed and Fluid: Stability and Change in the Geography of the Internet. *Telecommunications Policy* 26 (7-8): 389-413.

Gorman, S.P. and Malecki, E.J. (1999) The networks of the Internet: an analysis of provider networks. *Telecommunications Policy* 24: 113-134.

Gorman, S.P. and McIntee, A. (2001) Making Sense of the Urban Spectrum: Valuation, Agglomeration and Location of Wireless Technologies. Submitted for publication.

Gorman, S.P. (2002) "Spatial Small Worlds: Impacts of the Internet's Evolving Structure" American Association of Geographers, Los Angeles, CA

Grubesic, T. H. , O'Kelly, M. E., and Murray, A.T. (Forthcoming) A geographic perspective on telecommunication network survivability. *Telematics and Informatics*.

Haggett, P. and Chorley, R. (1969). Network Analysis in Geography. New York: St. Martins Press.

Hayes, B. (2000a) Graph theory in practice: Part I. *American Scientist* 88 (01): 9-13.

Huberman, B. and Ademic L. (1999) Growth dynamics of the World Wide Web. *Nature* 401:131.

Kansky, K. (1963). *Structure of Transportation Networks: Relationships Between Network Geometry and Regional Characteristics.* University of Chicago,
Department of Geography, Research Papers.

Kunreuther, H., Heal, G. and Orszag, P. (2002) Interdependent Security: Implications for Homeland Security Policy and Other Areas. *The Brookings Institute* Policy Brief #108.

Longcore, T. and Rees, P. (1996) Information Technology and Downtown Restructuring: The Case of New York City's Financial District. *Urban Geography* 17: 354-372.

Lowe, J. and Moryadas, S. (1975) *The Geography of Movement*. Prospect Heights,

IL: Waveland Press.

Malecki, E.J. (2001) The Internet Age: Not the End of Geography, in D. Felsenstein and M.J. Taylor, eds. *Promoting Local Growth: Process, Practice and Policy.* Aldershot: Ashgate, 2001, pp. 227-253.

Malecki, E.J. and Gorman, S.P. (2001) Maybe the death of distance, but not the end of geography: the Internet as a network, in S.D. Brunn and T.R. Leinbach (eds.) *The Worlds of Electronic Commerce*. New York: John Wiley, pp. 87-105.

Malecki, E.J. (2002) The Internet: A preliminary analysis of its evolving economic geography, *Economic Geography*, vol .78 , no. 4, pp. 399-424.

Medina, A., Matta, I., and Byers, J. (2000) On the origin of power-laws in Internet topologies. *ACM Computer Communications Review* 30 (2).

Milgram, S. (1977) The small world problem. In *The Individual in a Social World: Essays and Experiments*, pp. 281-295. Reading, Mass.: Addison-Wesley.

Moss, M. (1998) Technologies and Cities. *Cityscape: A Journal of Policy Development and Research* 3: 107-127.

Moss, M.L. and Townsend, A. (1997) Tracking the net: using domain names to measure the growth of the Internet in US cities. *Journal of Urban technology* 4 (3): 47-60.

Moss, M.L. and Townsend, A. (2000) The Internet backbone and the American metropolis. *The Information Society* 16: 35-47.

Negroponte, N. (1995) *Being Digital*. New York: Alfred A. Knopf.

NRC (2002) *Cybersecurity Today and Tomorrow: Pay Now or Pay Later.* Washington, DC: National Academy Press.

NSTAC (2002) ***Network Security/Vulnerability Assessments Task Force Report.*** Washington, DC: The President's National Security Telecommunications Advisory Committee. http://www.ncs.gov/nstac/NSVATF-Report-(FINAL).htm

O'Kelly, M. E. and Grubesic, T.H. (Forthcoming) Backbone topology, access, and the commercial Internet, 1997 – 2000. *Environment and Planning B.*

Odlyzko , A.M. (2001) Comments on the Larry Roberts and Caspian Networks study of Internet traffic growth, *The Cook Report on the Internet*, Dec., pp. 12-15.

Paltridge, S. (2002) Internet traffic exchange and the development of end-to-end international telecommunications competition. OECD: Working Paper
33

Pansiot, J. and Grad, D. (1998) On routes and multicast trees in the Internet. *ACM SIGCOMM Computer Communications Review* 28 (1): 41-50

Paxson, V. (1996) End-to-end routing behavior in the Internet. in *Proceedings of the ACM SIGCOMM 96'* Sept: 25-38.

Radoslavov, P., Tangmunarunkit, H., Yu, H. Govindan, r. Schenker, S. and Estrin, D. (2000) On characterizing network topologies and analyzing their impact on protocol design. *Tech Report 00-731*, University of Southern California, Dept. of CS.

Redner, S. (1998) How popular is your paper? An empirical study of the citation distribution. *European Physical Journal B* 4:131-134. http://xxx.lanl.gov/abs/cond-mat/9804163

Tangmunarunkit, H., Govindan, R., Jamin, S., Schenker, S., and Willinger, W. (2001) Network topologies, power laws, and hierarchy. Submitted for publication.

Telegeography (2002) *US Internet Geography 2002.* Washington DC: Telegeography Inc.

Townsend, A. (2001) Global cities, the Internet, and the flow of information". *American Behavioral Scientist*

Watts, D.J. and Strogatz, S.H. (1998) Collective dynamics of small-world networks. *Nature* 363: 202-204.

Wheeler, D.C. and O'Kelly, M.E. (1999) Network topology and city accessibility of the commercial Internet. *Professional Geographer* 51: 327-339.

Yasin, R. (2001) Gov't to map infrastructure. *Internetweek.com* http://www.internetweek.com/story/INW20011206S0001

Yook S.H., Jeong H., and Barabási, A.L. (2001) Modeling the Internet's Large-Scale Topology. http://xxx.lanl.gov/abs/cond-mat/0107417

**Table 1. Output of Diameter and S-I Index Analysis on Nodal Hierarchies**

| | Binary Hierarchy | | |
|---|---|---|---|
| **Diameter** | **CMSA** | I=u2/u1 | S=u3/u2 |
| 7 | | 0.2937 | 0.0499 |
| 8 | Atlanta | 0.3416 | 0.0927 |
| 8 | Chicago 1 | 0.3445 | 0.0449 |
| 8 | San Francisco 1 | 0.3466 | 0.0424 |
| 10 | Dallas 2 | 0.4415 | 0.4056 |
| 10 | Washington 1 | 0.4441 | 0.3019 |
| 10 | New York 1 | 0.4463 | 0.3133 |
| 10 | Denver 2 | 0.4602 | 0.3656 |
| 10 | Houston | 0.5313 | 0.4742 |
| 10 | Kansas City | 0.5410 | 0.3871 |
| 10 | Los Angeles | 0.5085 | 0.2671 |
| 10 | Cleveland | 0.5037 | 0.2268 |
| 10 | St. Louis 2 | 0.5096 | 0.1999 |
| 10 | Salt Lake City | 0.5069 | 0.1805 |
| 10 | Boston 2 | 0.5145 | 0.1185 |
| 10 | Phoenix | 0.5374 | 0.1309 |

| | Band Width Hierarchy | | |
|---|---|---|---|
| **Diameter** | **CMSA** | I=u2/u1 | S=u3/u2 |
| 7 | | 0.2937 | 0.0499 |
| 8 | New York 1 | 0.3029 | 0.0454 |
| 8 | Chicago 1 | 0.3063 | -0.0125 |
| 8 | San Francisco 1 | 0.3084 | -0.0135 |
| 9 | Dallas 2 | 0.4029 | 0.4381 |
| 9 | Washington 1 | 0.4086 | 0.3767 |
| 10 | Atlanta | 0.4442 | 0.2252 |
| 10 | Los Angeles | 0.4205 | 0.1234 |
| 10 | Seattle | 0.4188 | 0.1110 |
| 10 | Denver 2 | 0.4255 | 0.1322 |
| 10 | Kansas City | 0.4391 | 0.0860 |
| 10 | Salt Lake City | 0.4376 | 0.0664 |
| 10 | Houston | 0.4976 | 0.1549 |
| 10 | Boston 2 | 0.5530 | 0.1408 |
| 10 | Philadelphia | 0.5533 | 0.1429 |
| 10 | St. Louis 2 | 0.5624 | 0.1237 |

| | Small World Regional Hierarchy | | |
|---|---|---|---|
| **Diameter** | **CMSA** | I=u2/u1 | S=u3/u2 |
| 7 | | 0.2937 | 0.0499 |
| 8 | New York 1 | 0.3029 | 0.0454 |
| 8 | Chicago 1 | 0.3063 | -0.0125 |
| 8 | San Francisco 1 | 0.3155 | -0.0468 |
| 8 | Washington 1 | 0.3318 | -0.0793 |
| 9 | Boston 2 | 0.3938 | 0.2081 |
| 10 | Dallas 2 | 0.4804 | 0.4802 |
| 10 | Denver 2 | 0.4962 | 0.5025 |
| 10 | St. Louis 2 | 0.4982 | 0.4890 |
| 11 | Cleveland | 0.5915 | 0.6812 |
| 11 | Louisville | 0.5933 | 0.6759 |
| 11 | Kansas City | 0.6600 | 0.5959 |
| 12 | Seattle | 0.7778 | 0.9118 |
| 12 | Phoenix | 0.7752 | 0.8822 |
| 12 | Los Angeles | 0.7622 | 0.8810 |
| 12 | Atlanta | 0.7656 | 0.4362 |

| | Small World Distance Hierarchy | | |
|---|---|---|---|
| **Diameter** | **CMSA** | I=u2/u1 | S=u3/u2 |
| 7 | | 0.2937 | 0.0499 |
| 8 | Salt Lake City | 0.2935 | 0.0399 |
| 8 | Denver 2 | 0.3003 | 0.0573 |
| 8 | San Francisco 1 | 0.3061 | 0.0246 |
| 9 | Dallas 2 | 0.4081 | 0.5258 |
| 9 | Seattle | 0.4072 | 0.5149 |
| 9 | Chicago 1 | 0.4194 | 0.4465 |
| 9 | Los Angeles | 0.3841 | 0.2800 |
| 10 | Atlanta | 0.4205 | 0.1839 |
| 10 | Washington 1 | 0.4420 | 0.0249 |
| 10 | New York 1 | 0.4394 | -0.1134 |
| 10 | Phoenix | 0.4583 | -0.0784 |
| 11 | Houston | 0.5412 | 0.1520 |
| 13 | Miami | 0.7341 | 0.6719 |
| 14 | Boston 2 | 0.9572 | 0.8135 |
| 16 | Kansas City | 1.3219 | 1.1954 |

| Diameter | Global Hierarchy | | |
|---|---|---|---|
| | CMSA | I=u2/u1 | S=u3/u2 |
| 7 | | 0.2937 | 0.0499 |
| 8 | San Francisco 1 | 0.2981 | 0.0258 |
| 8 | Atlanta | 0.3489 | 0.0779 |
| 8 | Chicago 1 | 0.3518 | 0.0169 |
| 10 | Dallas 2 | 0.4384 | 0.3208 |
| 10 | Denver 2 | 0.4503 | 0.3691 |
| 10 | Washington 1 | 0.4717 | 0.1825 |
| 10 | New York 1 | 0.4672 | 0.0570 |
| 10 | Salt Lake City | 0.4649 | 0.0189 |
| 10 | Los Angeles | 0.4427 | -0.0806 |
| 10 | Houston | 0.4932 | -0.0264 |
| 11 | Kansas City | 0.5306 | -0.0190 |
| 11 | Seattle | 0.5317 | -0.0705 |
| 12 | Phoenix | 0.6464 | 0.2665 |
| 13 | Boston 2 | 0.8097 | 0.3425 |
| 16 | Miami | 1.3219 | 1.1954 |

| Diameter | Relay Node Hierarchy | | |
|---|---|---|---|
| | MSA | I=u2/u1 | S=u3/u2 |
| 7 | | 0.2937 | 0.0499 |
| 8 | Kansas City | 0.2958 | 0.0405 |
| 8 | Salt Lake City | 0.2956 | 0.0302 |
| 8 | Indianapolis | 0.2949 | 0.0227 |
| 8 | Seattle | 0.2942 | 0.0137 |
| 10 | Portland | 0.3654 | 0.6527 |
| 10 | Sacramento | 0.3834 | 0.7821 |
| 10 | St. Louis | 0.3866 | 0.7927 |
| 10 | Denver | 0.4063 | 0.7470 |
| 10 | Atlanta | 0.4248 | 0.5493 |
| 10 | Washington-Baltimore | 0.4254 | 0.4537 |
| 10 | Chicago | 0.4285 | 0.3020 |
| 10 | Philadelphia | 0.4291 | 0.2970 |
| 10 | Orlando | 0.4412 | 0.2912 |
| 12 | Jacksonville | 0.5249 | 0.6122 |
| 12 | Phoenix | 0.5237 | 0.6021 |

**Table 2. Link Frequency Analysis for the Genuity Fiber Network**

| | 2 hops | 3 hops | 4 hops | 5 hops | 6 hops | 7 hops | 8 hops | Total |
|---|---|---|---|---|---|---|---|---|
| **Denver - Kansas** | 5 | 11 | 18 | 23 | 34 | 34 | 23 | 148 |
| **El Paso - Houston** | 4 | 8 | 19 | 24 | 27 | 31 | 15 | 128 |
| **Both Cuts** | 9 | 19 | 37 | 47 | 61 | 65 | 38 | 276 |
| **Total Paths** | 60 | 79 | 95 | 97 | 86 | 67 | 38 | 522 |
| **% of Total** | **0.15** | **0.24** | **0.39** | **0.48** | **0.71** | **0.97** | **1.00** | **0.53** |

**Figure 1: Comparison of Exponential and Scale-Free Networks**

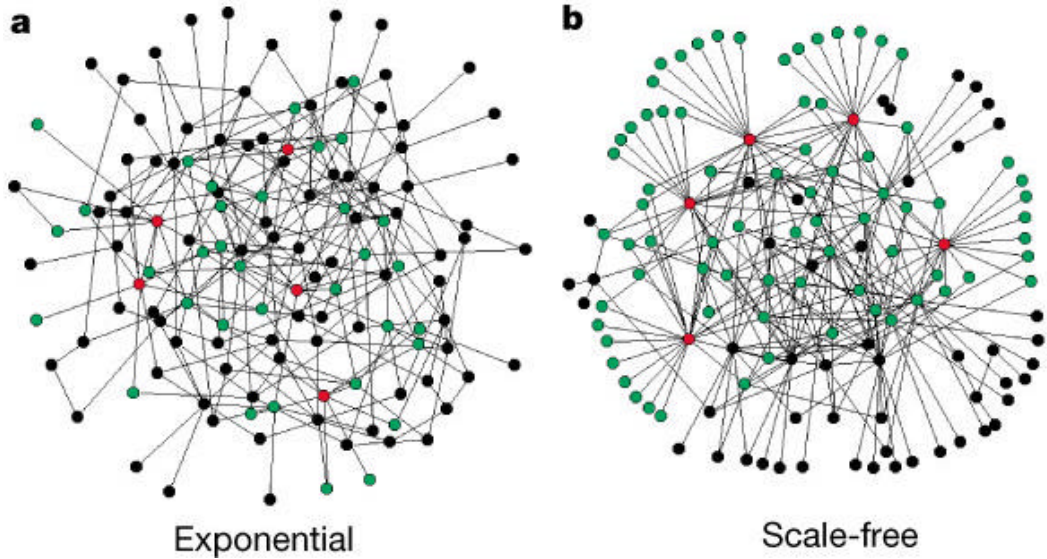

Note: Red dots are the five nodes with the highest number of links; green, their first neighbors. In the exponential network only 27% of the nodes are reached by the five most connected nodes, in the scale-free network more than 60% are reached, demonstrating the importance of the connected nodes in the scale-free network Both networks contain 130 nodes and 215 links ($k = 3.3$).

Source: Barabasi, A. (2001)

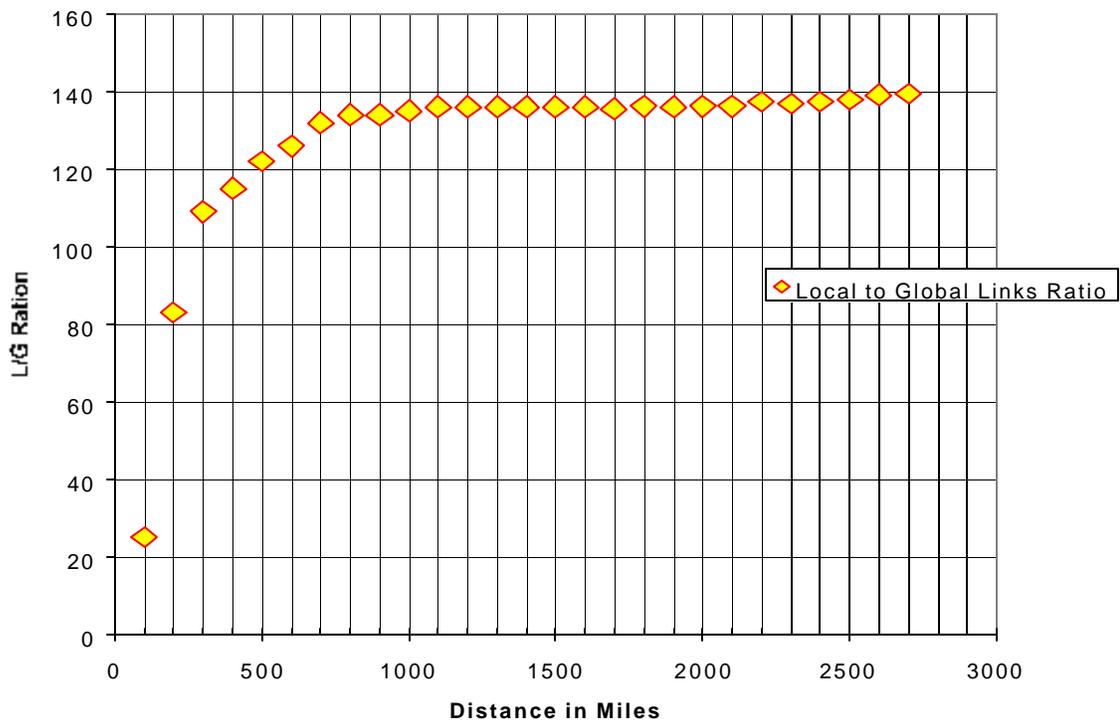

Fig 2. Local to Global Links Ratio by Distance

**Fig 3. Rate of Change of Local to Global Link Ratio**

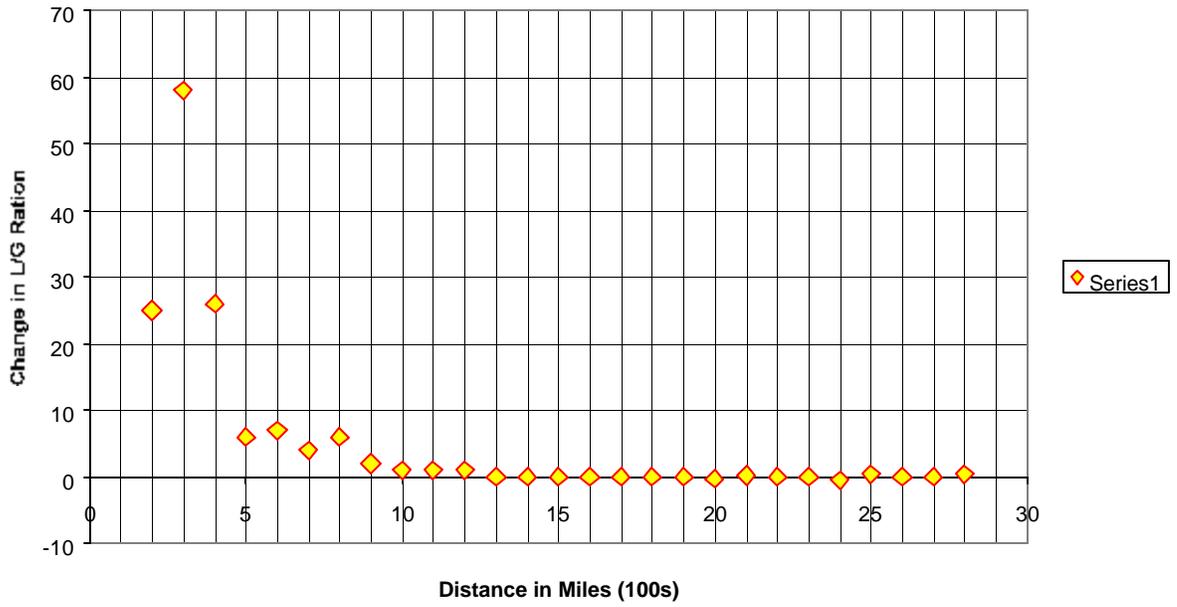

**Figure 4. Comparison of the Diameter Effects of Nodal Hierarchies**

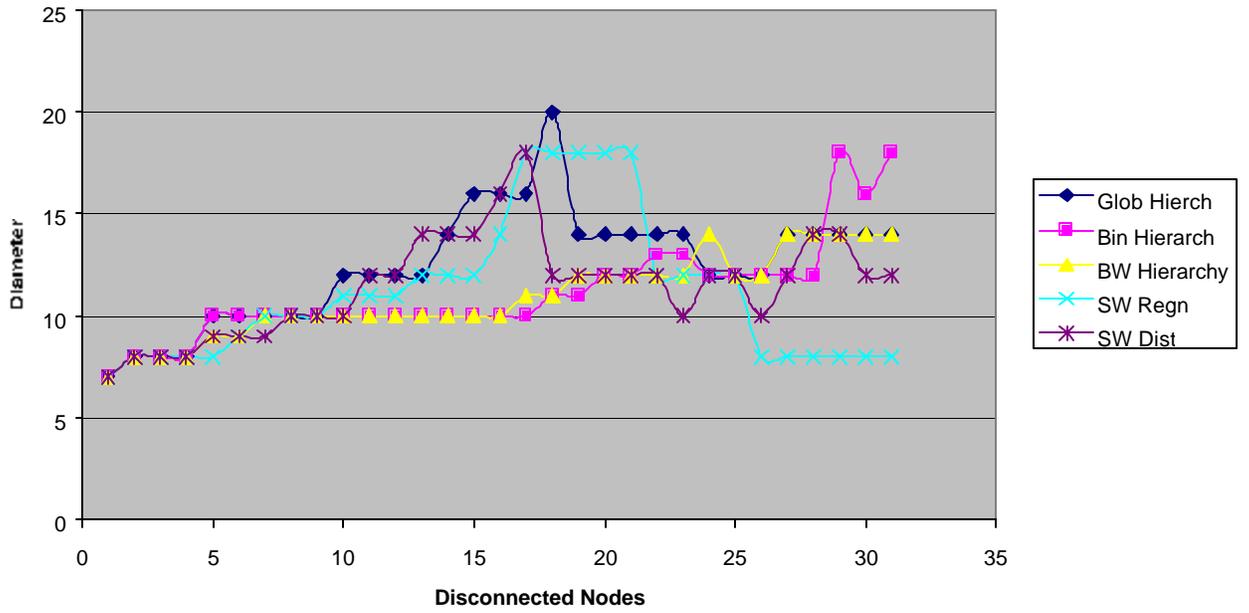

**Figure 5. Global Hierarchy: S and I Index and Diameter of the Network**

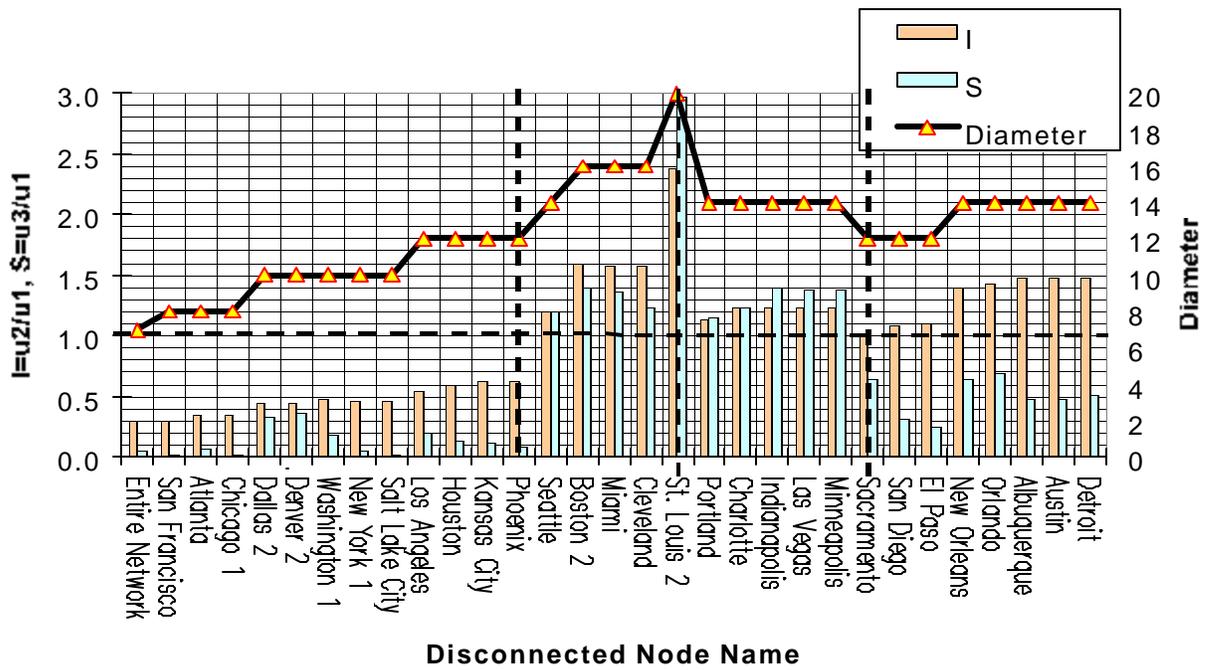

**Figure 6**

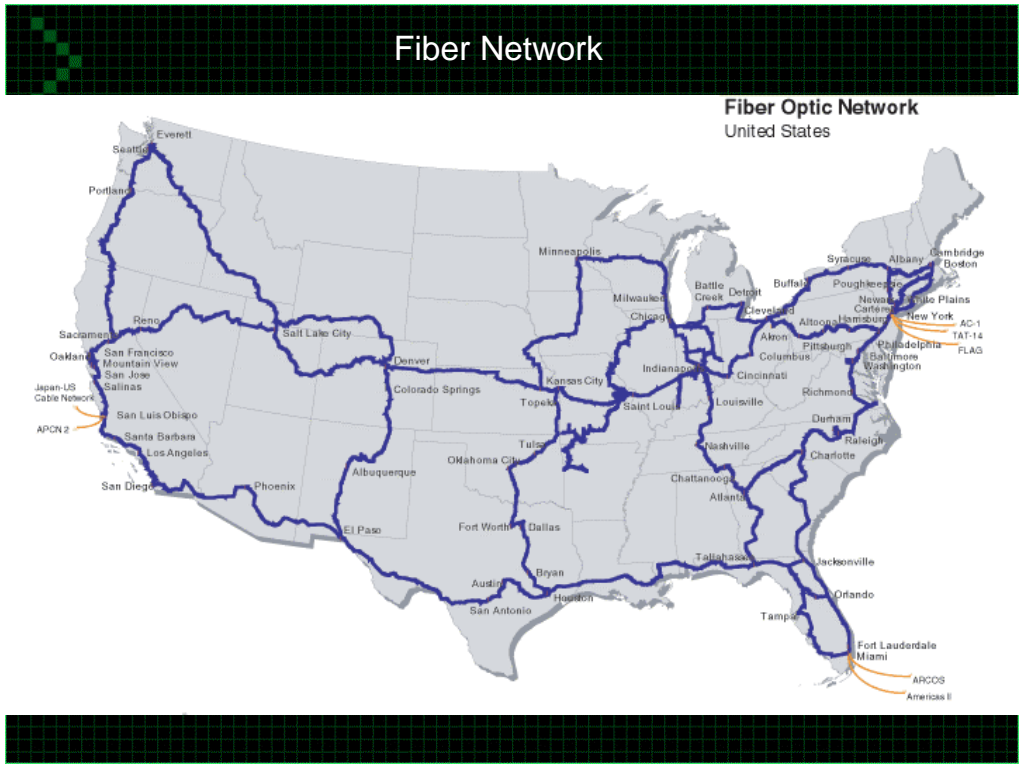

**Figure 7**

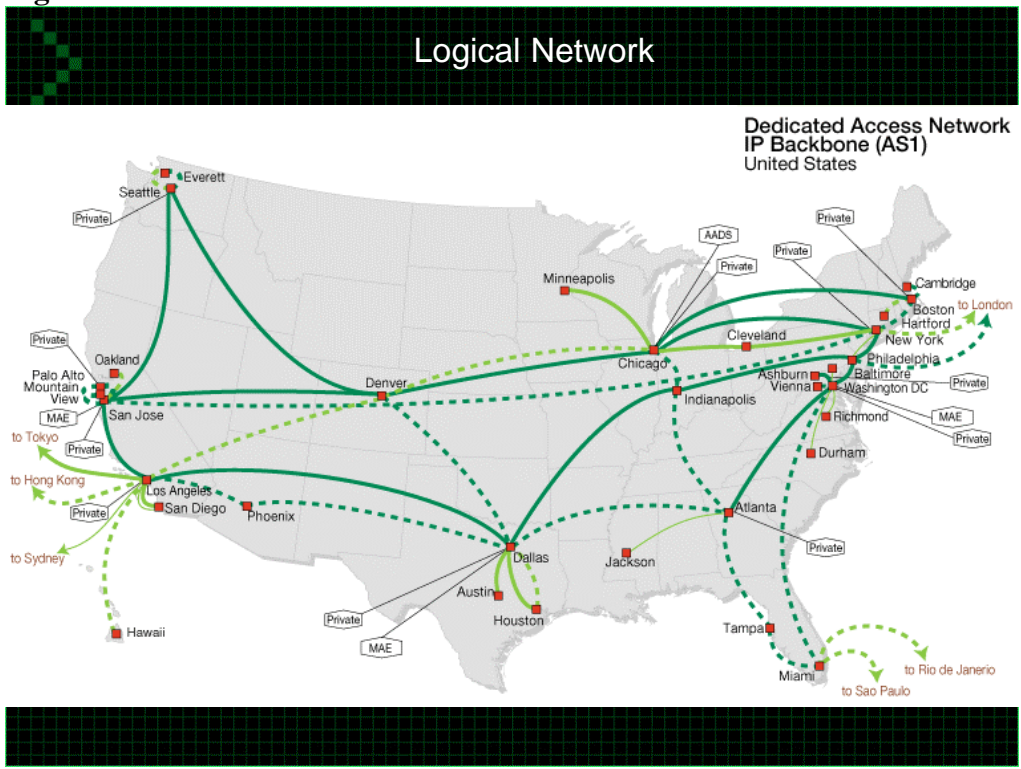

**Figure 8**